\documentclass[runningheads]{svmult}

\usepackage{makeidx}   
\usepackage{graphicx}  
\usepackage{subeqnar}  
\usepackage{multicol}  
\usepackage{physprbb}  
\makeindex             

\begin{document}
\title*{Planetary Nebulae in NGC~5128 (Centaurus~A)}
\toctitle{Planetary Nebulae in NGC~5128 (Centaurus~A)}
%
%
\titlerunning{PNe in NGC 5128}
%
\author{Eric W.\ Peng$^1$, Holland C.\ Ford$^2$, Kenneth C.\ Freeman$^3$}
\authorrunning{Eric Peng}
%
%
\institute{$^1$Rutgers University, Piscataway, NJ 08854, USA \\
$^2$Johns Hopkins University, Baltimore. MD 21218, USA \\
$^3$Australian National University, Canberra, ACT, Australia}
\maketitle              

\begin{abstract}
The study of planetary nebulae (PNe) in the nearby post-merger
elliptical galaxy NGC~5128 (Cen~A) now has a history of nearly twenty
years.  As the nearest giant elliptical, it is a prime target for
extragalactic PN studies.  These studies have addressed many issues
including the galaxy's distance, dark matter content, halo structure,
merger history, and stellar populations.  We review the main PN studies
that have been conducted in NGC~5128, and introduce a new study where we
measure the [NII]/H$\alpha$ ratio for 134~PNe.  We find that there are
no PNe in our sample that are obviously of Type~I, supporting the idea that the
last major star formation event in the galaxy halo occurred 
over 1--2~Gyr ago.
\end{abstract}

\section{Introduction}
The study of planetary nebulae (PNe) beyond the Local Group has largely been
conducted in luminous early-type galaxies where PNe are present in
large numbers and are relatively easy to detect.  The nearest, easily 
observable elliptical galaxy, NGC~5128 (Cen~A), has thus been a
recurring target for PN studies over the last twenty years.  At a
distance of 3.5--4~Mpc (e.g.\ Hui et al.\ 1993a; Harris et al.\ 1999; 
Rejkuba 2004), NGC~5128 is $\sim3$~mag closer
than the ellipticals in the Virgo cluster, which not only makes
intrinsically fainter PNe more accessible, it also greatly reduces
the problem of contamination from Ly$\alpha$ galaxies at $z=3.14$.  The
galaxy is also interesting in its own right as it is clearly the product
of a recent merger, exhibiting a warped central gas and dust disk,
stellar and HI shells, and both radio and x-ray jets.  A thorough review
of this galaxy is presented by Israel (1998).  Aside
from its intrinsic complexity, disadvantages of studying NGC~5128
include its large extent on the sky ($1'\sim1$~kpc), and its
relatively low Galactic latitude ($b=+19$). Nevertheless, modern
wide-field and multiplexing instrumentation have made studies both
practical and fruitful.  

\section{PN Surveys}

\begin{figure}[t]
\begin{center}
\includegraphics[width=.7\textwidth]{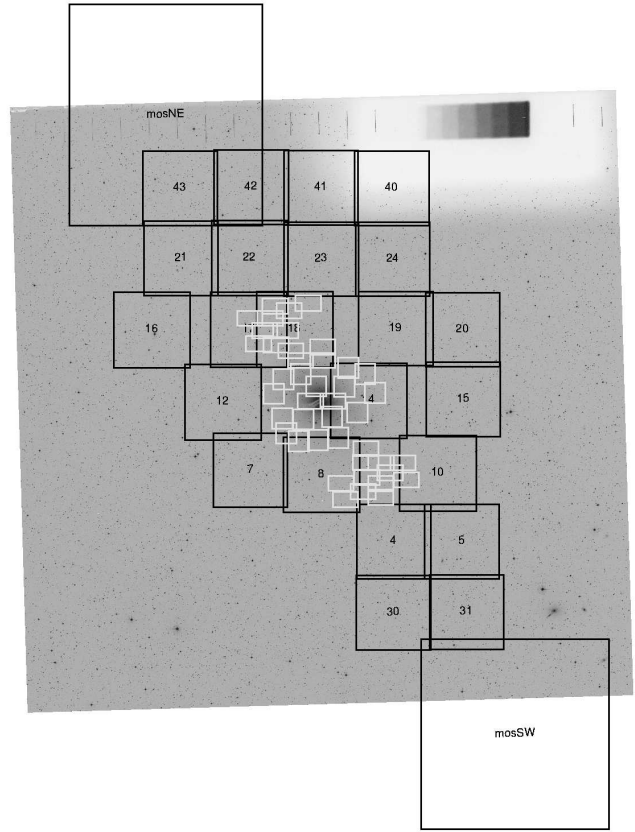}
\end{center}
\caption[]{NGC 5128 PN survey areas.  The smaller fields toward the
  center are the first survey fields by Hui et al.\ (1993a).  The larger
  fields that cover a much wider region of the halo are from our most
  recent survey (Peng et al.\ 2004a).}
\label{fig:fields}
\end{figure}

The first PN survey of NGC~5128 was conducted by Hui et al.\ (1993b),
and covered 661~arcmin$^2$ over 43 CCD fields (see Figure~\ref{fig:fields}).  
This coverage extended 20~kpc along the major axis an 10~kpc along the
minor axis.  At these projected distances ($\sim4$ and 2~$r_e$,
respectively) PNe were detected well beyond the radii at which long-slit
spectroscopy was readily obtainable.  They identified 785 PN candidates
using the method of on-/off-band imaging of the [OIII]$\lambda5007$
emission line.  These PNe formed a
complete sample for brightest 1.5~mag of the PN luminosity function
(PNLF).  They also spectroscopically confirmed 432 PNe, on which they
based their dynamical studies of dark matter and the intrinsic shape of
the potential (Hui et al.\ 1995).

Larger CCD cameras and fiber spectrographs allowed us to conduct a
second, wider PN survey of the NGC~5128 halo.  This survey covered
nearly 3~deg$^2$ on the sky and extended to 100~kpc along the major axis
and 40~kpc along the minor axis (Figure~1).  In total, we increased 
the number of imaged PNe to 1141, 780 of which are spectroscopically
confirmed.  The spatial distribution of PNe is concentrated toward the
major axis, mirrors that of the low surface brightness halo light seen in deep
photographic imaging by Malin (1978), although the most distant PNe in
our sample are $\sim30$~kpc beyond the light seen in the photographic
images.

\section{Mass Estimates}

\begin{figure}[t]
\begin{center}
\includegraphics[width=.7\textwidth]{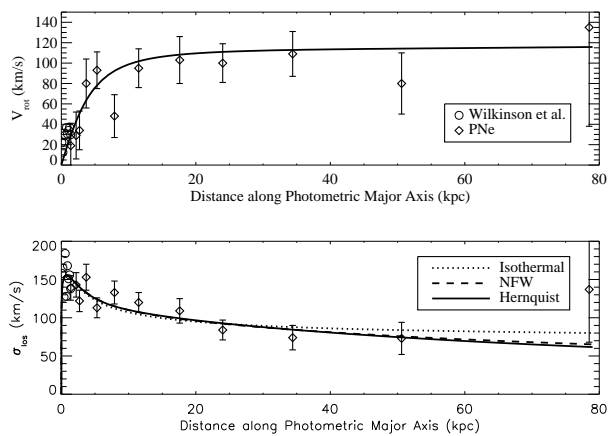}
\end{center}
\caption[]{(a) Major axis rotation curve.  We fit a parametrized flat
  rotation curve.  (b) Jeans Equation fit to the major axis
velocity dispersion profile.  In the bottom panel is the best fit
line-of-sight velocity dispersion profile for three different
dark halo mass models.}
\label{fig:rotsig}
\end{figure}

PNe are some of the most effective kinematic tracers in the halos of
elliptical galaxies, where their radial velocities can be used
to constrain the dynamical mass of their parent galaxies (e.g. Arnaboldi
et al.\ 1998; Romanowsky et al.\ 2003).  Hui et al.\ (1993) used the
original sample of 432 PN velocities to derive a total mass of
$3.1\times10^{11} M_{\odot}$ and $M/L_B\sim10$ out to a radius of 25~kpc.
Our recent work, which extends out to 80~kpc, shows that the total mass
increases only slowly with radius.  The fitted major axis rotation curve and
velocity dispersion profile are shown in Figure~\ref{fig:rotsig}.
Although the rotation curve remains flat, the dispersion profile falls
with radius.  Within this volume, we estimate a dynamical mass of
$\sim5\times10^{11} M_{\odot}$ and $M/L_B\sim13$ (Peng et al.\ 2004a).  
These numbers are
relatively unchanged whether we use the tracer mass estimator (Evans et
al.\ 2003) or spherical Jeans equation under the assumption of
isotropy.  While it may be that this value for $M/L_B$ can be produced
from theoretical expectations (see Napolitano, this volume), the value
we derive is certainly low, like the ones seen in the sample of
Romanowsky et al.\ (2003), when compared to luminous ellipticals such as
M87 and M49 (C{\^o}t{\'e} et al.\ 2001, 2003).

\section{The Two-dimensional Velocity Field}

\begin{figure}[t]
\begin{center}
\includegraphics[width=.9\textwidth]{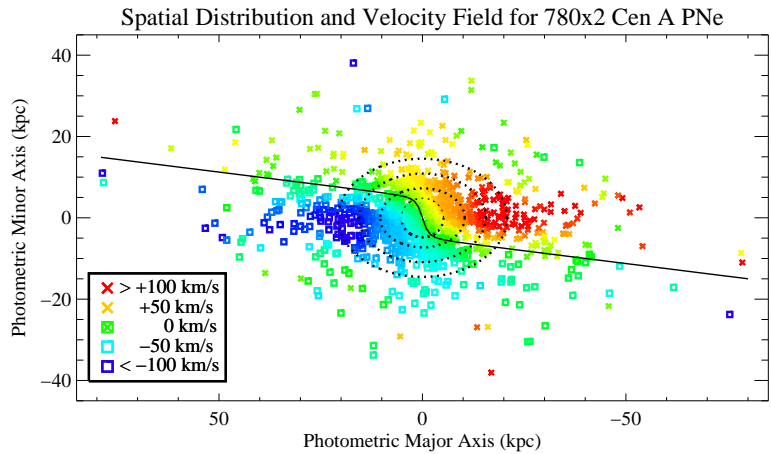}
\end{center}
\caption[]{Smoothed antisymmetrized velocity field with zero
  velocity contour (ZVC) plotted.  Dotted ellipses trace the rough
  isophotes for NGC~5128 from 1--4~$r_e$. PN with velocities larger than 
  systemic (receding) are represented by crosses, and those with velocities
  smaller than systemic (approaching) are represented by open boxes.  The
  color of each point shows the magnitude of the velocity with respect to
  systemic.  Notice how the
  rapid major axis rotation is only present within $\pm7$ degrees of the
  major axis, while the rest of the galaxy halo undergoes a slower minor
  axis rotation.}
\label{fig:vfield}
\end{figure}

One of the advantages of using PNe as kinematic probes is that we get
full two-dimensional spatial information on the stellar velocity field.  
When there are hundreds of measured PNe, we can construct velocity maps
that are similar to those derived from integral field spectrograph
observations, but that extend much farther out into the halo.  In the
case of NGC~5128, Hui et al.\ (1995) determined that the kinematic axis
of rotation was not aligned with either of the photometric axes,
suggesting that the intrinsic shape of the potential is triaxial.  Since
then, we have used our more extensive PN data to show that the
zero-velocity surface not only does not coincide with the photometric 
axes, but is in fact twisted in an 'S'-shape (Figure~\ref{fig:vfield}).  
This severe twist, which
asymptotes to a minor axis angle of 83~degrees, suggests that the
potential could be triaxial, tending toward prolate (Statler 1991).

Separate lines of evidence, which include our survey of globular
clusters in this system (Peng et al.\ 2004b) indicate that there may be
a substantial intermediate age population of stars with an mean age of 
$\sim5$~Gyr (Peng et al.\ 2004c).  Bekki et al.\ (2003) have already
produced simulations of globular cluster velocity fields that result
from galaxy mergers.  In the future, it will be valuable to
compare the PN velocity field to those produced by merger simulations to
constrain the types of mergers that could have produced this galaxy.

\section{PNe as Stellar Populations}

One of the useful characteristics of PNe is that, like globular
clusters, it is possible in priniple
to obtain both their radial velocities {\it and} some
information on their ages or abundances, producing a chemodynamical
picture of the galaxy.  The metallicity distribution of the
field star population of NGC~5128 has been measured using Hubble Space
Telescope color-magnitude diagrams of the red giants and is generally
metal-rich (Harris \& Harris 2002).  Assuming that the PNe trace the
same stellar population, this is one way to derive the overall
properties of the PN system.  However, one can derive abundances from PNe
directly from their emission line spectra, although this requires a
significant investment in observing time.  

Through a heroic effort, Walsh et al.\ (1999) derived [O/H] for 5 PNe in
NGC~5128, finding that they were generally sub-solar in abundance with a
mean value of $\sim-0.5$.  Walsh (this volume) has also recently
obtained spectra of much higher quality for $\sim50$ PNe, showing that
these types of observations are now possible and interesting.

Another way to link PNe with their progenitor stellar populations is to
find PNe in globular clusters.  Minniti \& Rejkuba (2002) found one such
object in an NGC~5128 globular cluster.  However, these objects appear
to be rare, and no other PNe in extragalactic clusters have been
reported.

We have investigated the PN population in NGC~5128 by looking for the
young, Peimbert (1978) Type~I PNe, which should have progenitors that
are younger than 1--2~Gyr.  In the Milky Way, these PNe are found in the
disk, and are rich in N and He due to the dredge-up and hot-bottom
burning processes found only in these more massive stars.  
Because obtaining spectra of
sufficient quality to derive N or He abundances would have been
prohibitively expensive, we instead chose to measure the [NII] lines
$\lambda\lambda6548,6584$ and compare their fluxes to H$\alpha$.  This
[NII]/H$\alpha$ ratio can be used as a crude estimator of relative
nitrogen abundance.  Type~I PNe in both Magellanic Clouds have very high
[NII]/H$\alpha$ ratios.

We used the 2-degree Field spectrograph on the Anglo-Australian
Telescope to observe 221 PNe in NGC~5128.  Of these, we detected
H$\alpha$ at $>3$-sigma in nearly all of them (219), and both H$\alpha$
and [NII]$\lambda6584$ in 134 PNe.  Because these were fiber
observations and some of the PNe sit atop the bright galaxy continuum,
it was also important that we model the galaxy and subtract it off.  In
order to do this, we used the models of Bruzual \& Charlot (2003) to
obtain a best fit simple stellar population to the blue side of the
spectrum (used in Peng et al.\ 2004b).  We then scaled this model to the
appropriate flux level, and shifted it to the expected mean velocity at
that position, as derived from our 2-d PNe velocity field.  Only after this
continuum subtraction did we measure the line fluxes.

\begin{figure}[t]
\begin{center}
\includegraphics[width=.80\textwidth]{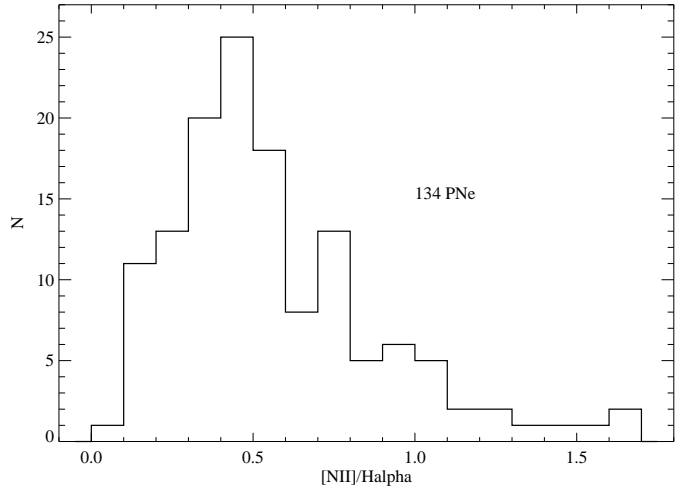}
\end{center}
\caption[]{Distribution of measured [NII]/H$\alpha$.  Our sample is
  complete for all values greater than $\sim0.55$.}
\label{fig:n2hahist}
\end{figure}

\begin{figure}[t]
\begin{center}
\includegraphics[width=.8\textwidth]{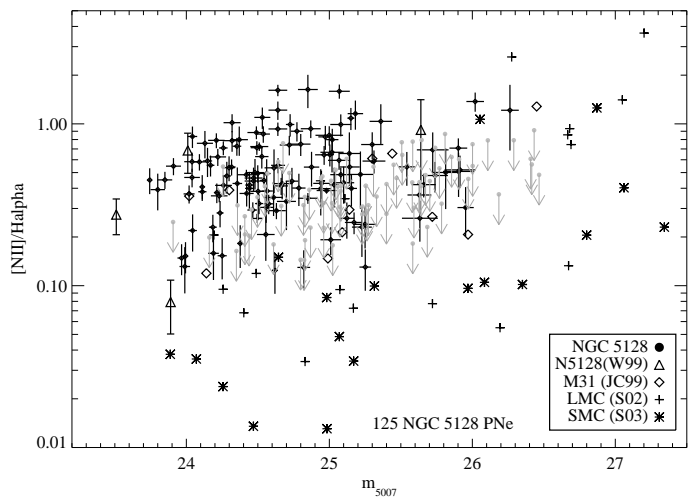}
\end{center}
\caption[]{[NII]/H$\alpha$ versus m$_{5007}$.  NGC~5128 values (points
  with error bars) show no evidence for obvious Type~I PNe.  Gray points
  are those for which we can only derive an upper limit for [NII/H$\alpha$.}
\label{fig:n2hamo3_log}
\end{figure}

Figure~\ref{fig:n2hahist} shows the distribution of measured
[NII]/H$\alpha$ values, with incompleteness setting in for values less
than $\sim0.55$.  In order to determine whether any PNe at the high end
of the distribution could plausibly be of Type~I, we compared them to
the extragalactic PNe found in the Magellanic Clouds (Stanghellini et
al.\ 2002, 2003), M31 (Jacoby \& Ciardullo 1999), and NGC~5128 (Walsh et
al.\ 1999).  In Figure~\ref{fig:n2hamo3_log}, we plot the value of
[NII]/H$\alpha$ against the brightness in [OIII]$\lambda5007$.  A
general trend in metallicity is evident with the SMC being most
metal-poor, and NGC~5128 being most metal-rich.  It is known that the 
Type~I PNe in the Clouds are typically faint, and this is seen
here, where the objects that have high [NII]/H$\alpha$ (relative to the
galaxy mean) are all at the faint end.  These
PNe can have [NII]/H$\alpha$ values that are $\sim10$~times higher than
what is typical in these galaxies.  In NGC~5128, however, while the
overall population has fairly high [NII]/H$\alpha$ values, we see no
evidence for outliers that would be good candidate Type~I PNe.  Even the
PN from Walsh et al.\ (1999) that was tentatively determined to be
Type~I, does not stand out from the crowd.
It is important to note that the
cutoff values of N/O and He/H typically used to define Type~I PNe are
defined only for the Galaxy, and may not apply to systems of different
metallicities.

\section{Conclusions}

Only in the Milky Way and M31 are there more known PNe than in
NGC~5128.  Starting with Hui et al.\ (1993a), extragalactic PN studies
in NGC~5128 have addressed a wide array of issues from its distance,
dark matter content, stellar populations, and formation history.  We now
know that while the halo of this galaxy does contain dark
matter, it falls into a growing class of elliptical galaxies that have
low to moderate mass-to-light ratios.  It's 2-dimensional halo
kinematics are consistent with a triaxial potential, and a galaxy that
was formed in a recent merger.  However, this merger was likely 
not so recent as to have produced many stars younger than $\sim2$~Gyr
that would now be detectable as Type~I PNe.  Future deep, wide-field
imaging and spectroscopy will be able to probe the PN abundance
distribution and the star formation history of this galaxy, as well as
enable studies of the PN stage of stellar evolution in an entirely
different environment from that in the Local Group.

%


\begin{thebibliography}{8.}
\addcontentsline{toc}{section}{References}

\bibitem{1998ApJ...507..759A} Arnaboldi, M., 
Freeman, K.~C., Gerhard, O., Matthias, M., Kudritzki, R.~P., M{\' e}ndez, 

\bibitem{2003MNRAS.338..587B} Bekki, K., 
Harris, W.~E., \& Harris, G.~L.~H.\ 2003, MNRAS, 338, 587

\bibitem{2003MNRAS.344.1000B} Bruzual, G.~\& 
Charlot, S.\ 2003, MNRAS, 344, 1000

\bibitem{2001ApJ...559..828C} C{\^ o}t{\' e}, P.~et al.\ 2001, ApJ, 559, 828 

\bibitem{2003ApJ...591..850C} C{\^ o}t{\' e}, P., McLaughlin, 
D.~E., Cohen, J.~G., \& Blakeslee, J.~P.\ 2003, ApJ, 591, 850 

\bibitem{1999AJ....117..855H} Harris, 
G.~L.~H., Harris, W.~E., \& Poole, G.~B.\ 1999, AJ, 117, 855 

\bibitem{2002AJ....123.3108H} Harris, W.~E.~\& 
Harris, G.~L.~H.\ 2002, AJ, 123, 3108

\bibitem{hui93} Hui, X., Ford,
  H.~C., Ciardullo, R., \& Jacoby, G.~H.\ 1993a, ApJ, 414, 463

\bibitem{1993ApJS...88..423H} Hui, 
X., Ford, H.~C., Ciardullo, R., \& Jacoby, G.~H.\ 1993b, ApJs, 88, 423

\bibitem{hui95} Hui, X., Ford, H.~C., Freeman, K.~C., \& Dopita, M.~A.\ 1995, ApJ, 449, 592

\bibitem{1998A&ARv...8..237I} Israel, F.~P.\ 1998, A\&A
  Rev., 8, 237 

\bibitem[Jacoby \& Ciardullo(1999)]{1999ApJ...515..169J} Jacoby, G.~H.~\& 
Ciardullo, R.\ 1999, ApJ, 515, 169

\bibitem{1978Natur.276..591M} Malin, D.~F.\ 1978, Nature, 276, 
591 

\bibitem{2002ApJ...575L..59M} Minniti, D.~\& 
Rejkuba, M.\ 2002, ApJL, 575, L59

\bibitem{1978IAUS...76..215P} Peimbert, M.\ 1978, IAU 
Symp.~ 76: Planetary Nebulae, 76, 215

\bibitem{2004ApJ...602..685P} Peng, E.~W., 
Ford, H.~C., \& Freeman, K.~C.\ 2004, ApJ, 602, 685

\bibitem{2004ApJS...150..367P} Peng, 
E.~W., Ford, H.~C., \& Freeman, K.~C.\ 2004b, ApJs, 150, 367

\bibitem{2004ApJ...602..705P} Peng, 
E.~W., Ford, H.~C., \& Freeman, K.~C.\ 2004c, ApJ, 602, 705

\bibitem{2004A&A...413..903R} Rejkuba, M.\ 2004, A\&A, 413, 
903

\bibitem{2003Science} Romanowsky, A.~J., 
Douglas, N.~G., Arnaboldi, M., Kuijken, K., Merrifield, M.~R.,
Napolitano, N.~R., Capaccioli, M., Freeman, K.~C.\ 2003, Science, 301, 1696

\bibitem{2002ApJ...575..178S} Stanghellini, L., 
Shaw, R.~A., Mutchler, M., Palen, S., Balick, B., \& Blades, J.~C.\ 2002, 
ApJ, 575, 178

\bibitem{2003ApJ...596..997S} Stanghellini, L., 
Shaw, R.~A., Balick, B., Mutchler, M., Blades, J.~C., \& Villaver, E.\ 
2003, ApJ, 596, 997

\bibitem{1991AJ....102..882S} Statler, T.~S.\ 1991, AJ, 
102, 882 

\bibitem{1999A&A...346..753W} 
Walsh, J.~R., Walton, N.~A., Jacoby, G.~H., \& Peletier, R.~F.\ 1999, A\&A, 
346, 753

\end{thebibliography}
\end{document}